\documentclass[%
 reprint,
 amsmath,amssymb,
 aps,
]{revtex4-2}

\usepackage{graphicx}% Include figure files
\usepackage{dcolumn}% Align table columns on decimal point
\usepackage{bm}% bold math
\usepackage[utf8]{inputenc}
\usepackage[T1]{fontenc}
\usepackage{mathptmx}
\usepackage[T1]{fontenc} % Use modern font encodings
\usepackage{amsmath,color}
\usepackage{amssymb}
\usepackage{amsfonts}
\usepackage{amsthm}
\usepackage{mathtools}
\usepackage[title]{appendix}
\usepackage{hyperref}
\usepackage{braket}
\usepackage{xcolor}
\usepackage{etoolbox}
\usepackage{soul}

\usepackage{algorithm}
\usepackage{algpseudocode}
\usepackage{dsfont}

\usepackage{bbold}

\renewcommand\thefigure{\arabic{figure}}
\begin{document}

\title{Quantifying the Chirality of Vibrational Modes in Helical Molecular Chains}
% Force line breaks with \\

\author{Ethan Abraham}
\email{abrahame@sas.upenn.edu}
\affiliation{Department of Physics and Astronomy, University of Pennsylvania, Philadelphia, Pennsylvania 19104, USA }
\author{Abraham Nitzan}
\affiliation{Department of Chemistry, University of Pennsylvania, Philadelphia, Pennsylvania 19104, USA }
\affiliation{School of Chemistry, Tel Aviv University, Tel Aviv 69978, Israel}

\date{\today}% It is always \today, today,
             
\begin{abstract}
\vspace{1 cm}Chiral phonons have been proposed to be involved in various physical phenomena, yet the chirality of molecular normal modes has not been well defined mathematically. Here we examine two approaches for assigning and quantifying the chirality of molecular
normal modes in double-helical molecular wires with various levels of twist. First, associating with each normal mode a structure obtained by imposing the corresponding motion on a common origin, we apply the Continuous Chirality Measure (CCM) to quantitatively assess the relationship between the chirality-weighted normal mode spectrum and the chirality of the underlying molecular structure. We find that increasing the amount of twist in the double helix shifts the mean normal mode CCM to drastically higher values, implying that the chirality of molecular
normal modes is strongly correlated with that of the underlying molecular structure. Second, we assign to each normal mode a pseudoscalar defined as the product of atomic linear and angular momentum summed over all atoms, and we analyze the handedness of the normal mode spectrum with respect to this quantity. We find that twisting the double-chain structure introduces asymmetry between right and left-handed normal modes so that in twisted structures different frequency bands are characterized by distinct handedness. This may give rise to global phenomena such as thermal chirality.

\end{abstract}
 \maketitle

Among the concepts with such far-reaching consequences across science, perhaps none has remained as quantitatively enigmatic as \textit{chirality}, defined as the non-superimposability of mirror images. While not impeding our understanding of its important consequences such as optical activity \cite{Barron_reference,optical_chirality}, enantioselective catalysis \cite{bio_enantioselectivity,enantioselective_transition_metal}, or homochirality of biological systems \cite{origin_homochir,DNA_Chiral_Recognition}, some important aspects of molecular chirality are not well understood. For example, it has been argued that there is no general, physically unambiguous method to assign handedness to molecular structures \cite{Kamien}. Furthermore, there has been no consensus on the mechanism of \textit{Chirality Induced Spin Selectivity} (CISS) \cite{CISS_overview, CISS_theory}, a phenomenon wherein electron transmission through and accumulation in chiral materials exhibits strong spin selectivity. 

The recently growing discussion of \textit{chiral phonons} illustrates the importance and difficulty of rigorously addressing chirality in condensed matter. Over the past half decade, both local \cite{cphonons_2D} and propagating \cite{cphonons_3D} chiral phonons have been proposed to play important roles in materials' electromagnetic phenomena, including angular
momentum transfer from photon to electron spin \cite{truly_chiral}, induction of strong magnetic fields by optically driven chiral phonons \cite{giant_magnetic}, spin selective transport \cite{CISS_phonon1, CISS_phonon2, CISS_phonon3}, and calorimetric phenomena such as spin Seebeck effect
\cite{Spin-Seedback}. 

Yet, the chiral phonon is still ill-defined, often leading to ambiguities. The challenge is that motions in molecules and condensed matter must be compositions of excited normal vibrational modes, which are material-specific and can possess non-trivial properties that undermine analogies with circularly polarized light (CPL) \cite{optical_chirality}. Phonon chirality is
sometimes identified with angular momentum \cite{cphonons_2D,cphonons_3D}; however it is not obvious that angular momentum implies chirality. Unlike in molecular structure where chirality is better defined, here the temporal dimension is at play, and Barron \cite{true_false_chir,Barron_reference} has pointed out that "true" chirality, where enantiomers are related by spatial inversion ($\mathcal{P}$), must be distinguished from "false" chirality, where enantiomers are related by time inversion ($\mathcal{T}$) followed by a spatial rotation ($\mathcal{R}$). Present descriptions of chiral phonons have relied on the symmetry properties and dispersion relations of crystal structures \cite{cphonons_3D,truly_chiral}, but these approaches are not directly applicable to the analysis of molecular normal modes. Furthermore, the connection between the chirality of such modes and the chirality of the underlying molecular structure is not well understood, although recent work indicates that such correlation does exist \cite{Chir_Phon_Chir_Cryst}. 

Mathematical methods for quantifying chirality, such as the \textit{Continuous Chirality Measure} (CCM) developed by Avnir \textit{et al.} \cite{c2,c3}, have recently been proposed and are now being applied in a variety of contexts. Alternatively, methods of quantifying chirality can be inferred from other fields in physics. The pseudoscalars $h=\textit{\textbf{p}}\cdot\sigma/|p|$ ($\textit{\textbf{p}}$ momentum; $\sigma$ spin) and $h=\textit{\textbf{v}}\cdot\omega$ ($\textit{\textbf{v}}$ velocity; $\omega=\nabla\times\textit{\textbf{v}}$ vorticity) are used to
characterize helicity in spintronics and fluid flow respectively. \cite{chiral_charge}. In optics, Tang and Cohen recognized that a chiral physical observable should be a time-even pseudoscalar \cite{optical_chirality}, which led them to define the chiral density of a field \textbf{A} as \begin{equation}\tag{1}\label{Cfield}C=\textbf{A}\cdot\nabla\times\textbf{A}.\end{equation} Indeed, the sum $(C=\frac{\epsilon_0}{2}\textbf{E}\cdot\nabla\times\textbf{E}+\frac{1}{2\mu_0}\textbf{B}\cdot\nabla\times\textbf{B}$) has been shown to determine the magnitude of circular dichroism \cite{optical_chirality,CD_analysis} in isotropic chiral samples. Similarly, when three ordered vectors $(\textbf{v}_1,\textbf{v}_2,\textbf{v}_3$) characterize an object, it has been suggested that chirality be defined by the scalar triple product $\textbf{v}_1\cdot(\textbf{v}_2\times\textbf{v}_3)$ \cite{chiral_fermions, chiral_fermions2, chiral_charge, orig_triple_prod, chiral_momentum, tools_benchmarks} \footnote{Note for molecular systems, a variation of the scalar triple product method has been proposed summing the triple-products $\Sigma_{i}(\textbf{v}_i\times\textbf{v}_{i+1})\cdot\textbf{v}_{i+2}$ of successive separation vectors connecting reference points ($\textbf{v}_i = \textbf{r}_{i+1}-\textbf{r}_{i})$. This method has been shown useful when the reference points are corresponding locations on successive residues in biological proteins \cite{bio_chir_char, dipole_chir_char}. We choose to address this triple product method in a separate work. \cite{tools_benchmarks}}. Apart from the CCM, these measures are all pseudoscalars composed of the inner product of a vector and a pseudovector, and they all appeal to our intuition regarding a helical structure defined by circulation about a central axis with a component parallel to that axis. 

In the present work we introduce quantitative procedures to characterize the chirality of molecular vibrational modes. Furthermore, we examine the correlation between the calculated normal mode chirality measure and the chirality of the underlying molecular equilibrium structure. As discussed below, our approach is different from the recent literature which associates chiral phonons with global excitations of chiral trajectories \cite{phon_ang_mom,cphonons_3D}; here we look at the geometry of individual normal modes rather than the collective motions they comprise.

This study builds on our recent work \cite{twist_paper}, which used MD simulations to model a polyethylene double-helical wire with various levels of (left-handed) twist. Here, we use harmonic analysis of these same structures to quantitatively explore the chirality of individual normal modes and their relationship to the chirality of the underlying molecular structure. The two-stranded polymer is an excellent model system for such a study because it can vary continuously between an achiral (untwisted) form and highly chiral (twisted) form. This allows us to compare the chirality of normal modes to that of the underlying equilibrium molecular structure. The results shown in this main text use polymers of length $N=98$ modeled by the TraPPE United Atom (UA) force field, which coarse grains each CH$_x$ unit into a single interaction site \cite{twist_paper, FF3, FF5}. The Supplemental Material \footnote{See Supplemental Material [url], which
includes Refs. \cite{twist_paper,Hadi,FJC,FF3,FF5,unitedWorks}, for a description of the force fields examined in our study.} confirms that our main findings persist using other polymer lengths and force fields as well. We find that the normal mode spectra of twisted structures show strikingly more chiral features than the untwisted control.

{\it CCM of Static Structures and Vibrational Modes.}  Chirality, to be distinguished from helicity, is defined by the non-superimposability of mirror images, manifesting in the absence of mirror symmetry \cite{c1}. \textit{Continuous symmetry measures} \cite{c3} assess the overlap of an object with the most similar object that contains a particular symmetry. In the case of chirality, this means measuring the overlap of an object with its nearest achiral object. If $\ket{Q}$ is an object in a vector space $V$, then the nearest achiral object is $\ket{Q'}=\frac{1}{2}(\mathds{1}+\sigma)\ket{Q}$, where $\sigma$ is a mirror reflection operator chosen so that $\braket{Q|Q'}$ is maximized. Then the normalized quantity $\braket{Q|Q'}/\braket{Q|Q}$ is always unity for an achiral structure and approaches ½ for a very chiral structure (for an analytical method of finding this optimal mirror plane, see Ref. \cite{c3}, and note the requirement that the mirror plane pass through the origin). Therefore, $\braket{Q|Q'}/\braket{Q|Q}$ is a measure of the mirror symmetry content of $\ket{Q}.$ The corresponding CCM is defined as \cite{c2,c3} (since chirality concerns the \textit{absence} of mirror symmetry)
\begin{equation}\tag{2}CCM(Q) = 1 - \frac{\braket{Q|Q'}}{\braket{Q|Q}}.\end{equation}
This measure vanishes for an achiral object and increases up to ½ as the chiral character increases \footnote{Originally, some inventors of the CCM intended it to range from 0 to 1, but it now appears that for most applications the upper bound of the CCM as defined is in fact ½}. In a standard application of this measure, the $Q$ is a molecular structure represented by the mass-weighted atomic coordinates $\ket{\textbf{q}_i}=\sqrt{m_i}(x_i,y_i,z_i)^T$ defined relative to the center of mass, and its CCM can naturally \footnote{The numerator in the second term of Eq. (3) is often defined as $\text{max}_P\Sigma_i\bra{\textbf{q}_i}\mathds{1}+\sigma P\ket{\textbf{q}_i},$ where $P$ is a permutation $P\ket{\textbf{q}_i}=\ket{\textbf{q}_j}.$ In this text we set $P=\mathds{1}.$ This choice is suggested by the observations that for helical chains, the trivial permutation tends to yield the lowest structural CCM as desired.} be calculated according to \begin{equation}\tag{3}\label{ccm_mol} CCM(Q) = 1 - \frac{\sum_{i = 1}^{N} \bra{\textbf{q}_i}\mathds{1}+\sigma\ket{\textbf{q}_i}}{2\sum_{i = 1}^{N} \braket{\textbf{q}_i | \textbf{q}_i}},\end{equation} 
\begin{figure}[!hbt]
\includegraphics[width=1\columnwidth]{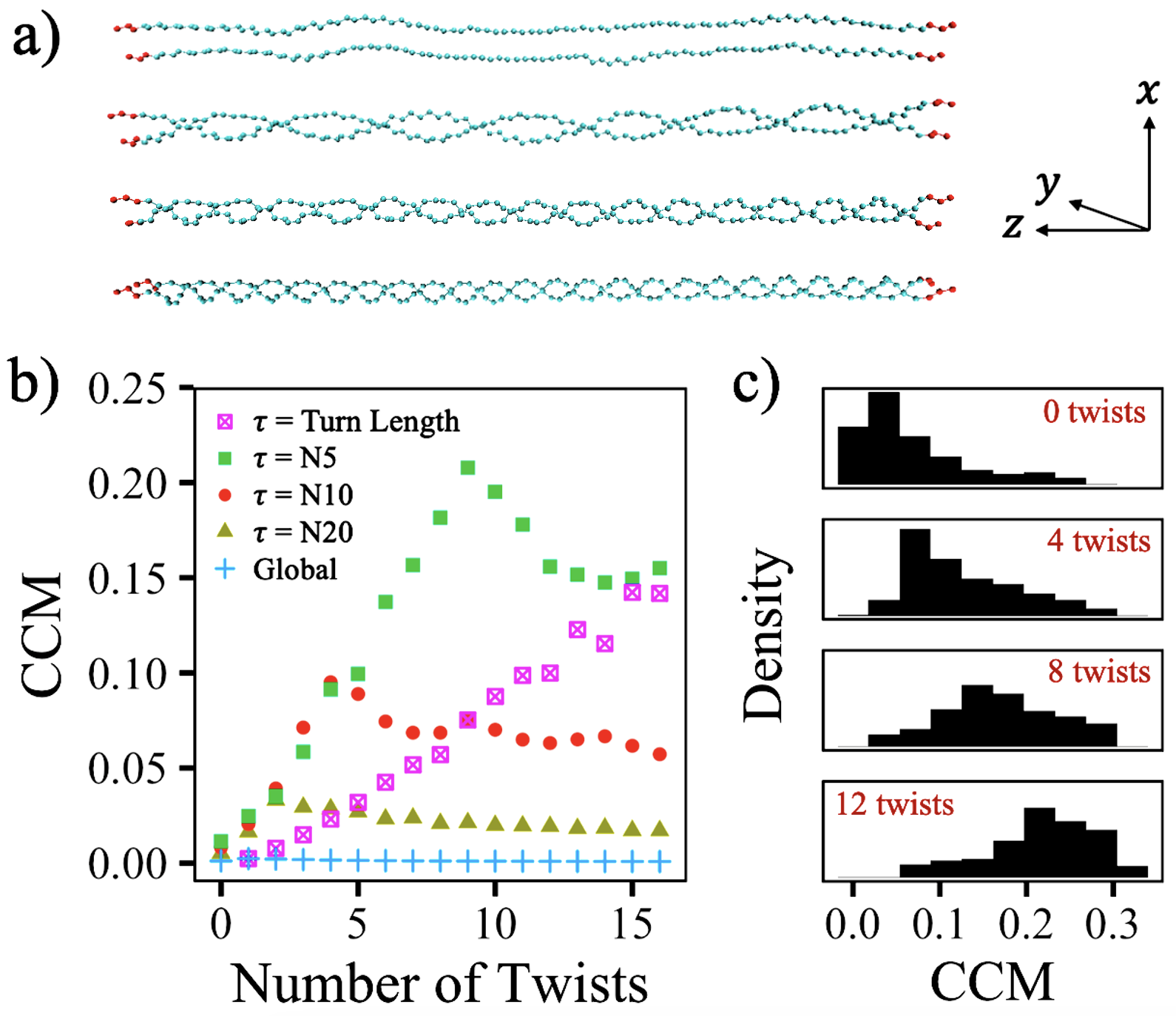}
\caption{(a) The model system for this study: a two-stranded $N=98$ polyethylene wire containing various levels of (left-handed) twist (shown from top to bottom: 0, 4, 8, 12 twists). The $z$-axis is taken to be the axis of the chain. Terminal atoms shown in red. (b) The CCM (Eq. (\ref{ccm_mol})) of the wire as a function of the number of twists. Results are shown when the structure $Q$ is taken to be a segment of the wire corresponding to (pink) one helical turn, (green, red, gold) 5, 10, or 20 monomers, and (blue) the entire structure. (c) Histograms of the normal mode distribution binned by CCM for the structures shown in the top panel.}
\label{FIG1}
\end{figure}where $\braket{\textbf{q}_i | \textbf{q}_i} = m_i(x_i^2+y_i^2+z_i^2)$ and $\braket{Q|Q} = \sum_i\braket{\textbf{q}_i | \textbf{q}_i}.$ Importantly, the same CCM concept may be applied to define the chirality of any molecular normal mode $k$ represented by atomic displacement vectors $\ket{\textbf{c}_{k,i}}=\sqrt{m_i}(c_{k,i}^x,c_{k,i}^y,c_{k,i}^z)^T$, evaluating the CCM by taking $\ket{\textbf{q}_i} \mapsto \ket{\textbf{c}_{k,i}}$ in Eq. (\ref{ccm_mol}). This makes it possible to assign a chirality measure to each normal mode, and in turn to evaluate the chirality-weighted molecular vibrational spectrum. Note that one weakness of the CCM measure (to be addressed later), is that it is not a pseudoscalar and therefore does not assign handedness to enantiomers: opposite enantiomers have the same CCM value.

In \hyperref[FIG1]{Fig. 1} we show the results of CCM calculations for both the equilibrium configurations of our two-stranded wires and the corresponding normal mode spectrum. \hyperref[FIG1]{Figure 1(a)} shows the model system for this study. \hyperref[FIG1]{Figure 1(b)} shows the CCM of the molecular structure as a function of the number of twists; note that when we choose $Q$ to be the entire wire (blue line), the CCM does not register twistedness. This is because the axial coordinates of the atoms $\ket{\textbf{q}_i}$ far from the center of mass dominate the inner products in Eq. (\ref{ccm_mol}), leading to low CCM values. When $Q$ is taken to be a segment of the wire on a shorter length scale $\tau$, the CCM is much more sensitive to the number of twists. Indeed, we have found that the most useful application of this concept is obtained when $\tau$ is taken to be a single helical pitch. With this choice, the CCM is roughly proportional to the twistedness of the double-helix. In the other cases when $\tau$ is a fixed length, a maximum is obtained when the number of twists is $\sim2\tau.$

\hyperref[FIG1]{Figure 1(c)} shows the first key result of this work, which is that the distribution of normal modes, as measured by the CCM, is much more chiral when the underlying structure is more twisted. Note that the CCM of the modes tends to be larger than the CCM of the structure \footnote{Note that that the majority of modes of the untwisted structure have a non-zero CCM; what changes with increased twist is the mean of the distribution. In this regard, it is worth noting that the untwisted structure showed in Fig. 1 may not be fully achiral because local minima of the untwisted wire's potential energy surface can be slightly asymmetric.}. Importantly, note that since the vectors that represent the normal modes are displacements from equilibrium rather than locations in extended space, the length scale considerations for \hyperref[FIG1]{Fig. 1(b)} are not relevant. Indeed, as shown in Fig. S1, the normal mode CCM distributions are qualitatively the same for an $N=36$ double helix as for the $N=98$. For the same reason, while the structural CCM depends on the choice of origin (typically taken to be the molecular center of mass), the CCM defined above for the normal modes does not.

{\it Momentum Pseudoscalar and Thermal Chirality.} The CCM shows a clear correlation between the chirality of normal modes and that of
the underlying molecular structure. Yet the physical meaning of the CCM is not transparent and, not being a pseudoscalar, it cannot be associated with handedness. In what follows, we consider physical quantities that can be associated with the chirality of molecular normal modes as well as their handedness.

First we must clarify our notation. An atomic Cartesian basis for deviations of $N$ atoms from their equilibrium positions is the collection of $3N$ vectors {$(0,\ldots,0,1,0,\ldots,0)^T,$ where consecutive triplets correspond to the three Cartesian displacements of a single atom. In this basis, a particular displacement of atoms from their equilibrium positions is written in terms of mass-weighted coordinates as $(m_1^{-1/2}\delta x_1,m_1^{-1/2}\delta y_1,m_1^{-1/2}\delta z_1,m_2^{-1/2}\delta x_2,\ldots)^T.$ In the same basis, a normal mode $k$ is written as $A_k\textbf{c}_k = A_k (c_{k,1}^x,c_{k,1}^y,c_{k,1}^z,c_{k,2}^x,\ldots)^T$ where the $\{\textbf{c}_k\}_{k=1...3N}$ constitute an orthonormal set ($\Sigma_i\Sigma_\alpha (c^\alpha_{k,i})^2 = 1$) and $A_k$ is the amplitude (of dimensionality $[\sqrt{m}l]$). When the normal mode $k$ has amplitude $A_k,$ the atomic displacement and velocity vectors are $A_k(m_1^{-1/2}c_{k,1}^x,m_1^{-1/2}c_{k,1}^y,m_2^{-1/2}c_{k,1}^z,m_2^{-1/2}c_{k,2}^x,\ldots)^T$ and $i\omega_kA_k(m_1^{-1/2}c_{k,1}^x,m_1^{-1/2}c_{k,1}^y,m_1^{-1/2}c_{k,1}^z,m_2^{-1/2}c_{k,2}^x,\ldots)^T,$ respectively (since $A_k(t) = A_k$exp$(i\omega_k t)$).

The association of chiral phonons with polarization in recent literature \cite{cphonons_2D, cphonons_3D, phon_ang_mom} is based on the modes' angular momentum. As discussed by Zhang and Niu \cite{phon_ang_mom}, the angular momentum associated with the atomic motions relative to their equilibrium positions is given by $\textbf{J} = \sum_i \textbf{u}_i \times \dot{\textbf{u}}_i$, where $\textbf{u}_i$ is the atomic displacement from equilibrium of atom $i$ and $\dot{\textbf{u}}_i$ is the corresponding atomic velocity. As we derive in the Supplemental Material \footnote{See Supplemental Material [url], which
includes Ref. \cite{phon_ang_mom}, for an analysis of phonon angular momentum.}, the axial angular momentum ($J^z$) of mode $k$ is proportional to $\sum_{i}\text{Im}[c_{k,i}^{x*} c_{k,i}^y].$ Since the normal mode coefficients $c_{k,i}^{\alpha}$ are real, it follows that the
angular momentum of a non-degenerate normal mode is zero, while for degenerate normal
modes we can construct linear combinations that will possess angular momentum. Such
angular-momentum carrying modes may be important in analyzing molecular response to
circularly polarized light, but the arbitrary choice of the linear combination leaves open the correspondence to the chirality of the underlying equilibrium structure.

Another option is to define angular momentum relative to some molecular reference frame \cite{phon_ang_mom}. For example, in our model the angular momentum of mode $k$ relative to the molecular axis is $L^z_k = \sum_i L_{k,i}^z = \dot{A}_k(t)\sum_i \sqrt{m_i}(x_i c^y_{k,i} - y_i c^x_{k,i}).$ However, note that under Eckart conditions this variable is strictly zero for an isolated molecule, and by itself it does not carry information about the molecular chirality since it is not a pseudoscalar.

\begin{figure}[!htb]
\includegraphics[width=1\columnwidth]{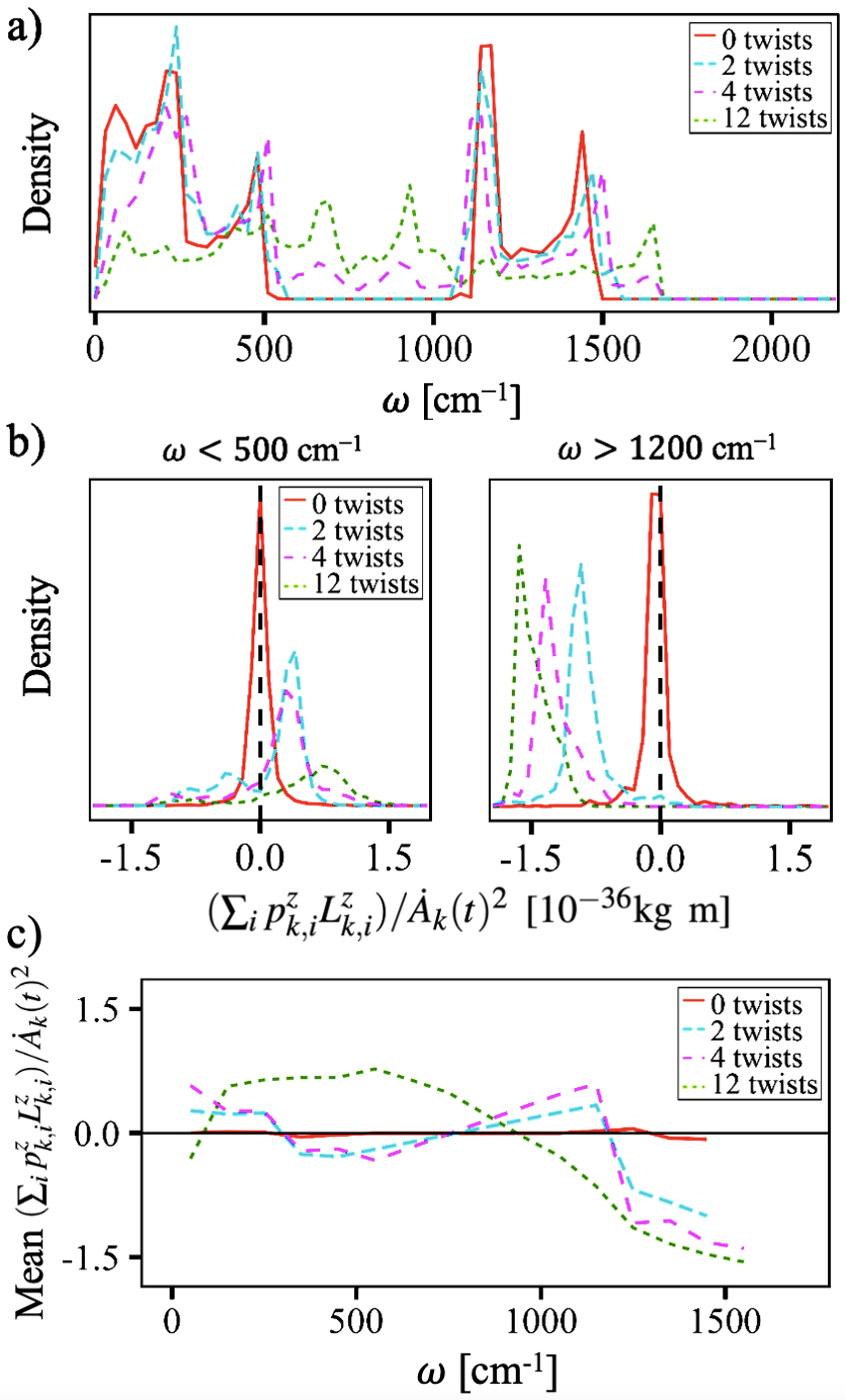}
\caption{(a) Frequency polygons showing the spectral densities of the two-stranded $N=98$ polyethylene wire when the wire is (red) untwisted, or (blue, purple, green) containing  2, 4, and 12 twists. All twists are left-handed. (b) Frequency polygons showing the density of modes binned by an axial momentum pseudoscalar (MPS) score (based on Eq. (\ref{PzLz})) for the same structures as the top panel. The left panel plots only the modes from the low frequency band, and the right from the high frequency band. (c) Spectrum of the MPS as a function of frequency obtained by averaging $\Sigma_i p_{k,i}^zL_{k,i}^z$ for modes within bin width $\Delta(\omega).$}
\label{FIG2}
\end{figure}
A better approach, which leads to the second key result of this work, is to look for a pseudoscalar in analogy with Eq. (\ref{Cfield}), taking $\textbf{A}\mapsto\textbf{p}_i,$ and $\nabla\times\textbf{A}\mapsto\textbf{L}_i.$} The resulting pseudoscalar $\textbf{p}_i\cdot\textbf{L}_i$ vanishes in general, and the components depend on the choice of origin. Nevertheless, many chirality-dependent physical processes take place along a particular axis (e.g. an electron's linear trajectory through a chiral material or a photon passing through a sample), so the essence of the behavior may be captured by $p_i^zL_i^z$ where the $z$-axis is the axis of interest (note that while $\textbf{p}_i \cdot \textbf{L}_i=0$, a Cartesian component $p_i^zL_i^z$ need not be zero). This provides a measure of the correlation between angular motion about the $z$-axis and linear momentum along this axis.

To apply this notion to a particular normal mode $k$, note that the oscillating linear momentum of atom $i$ moving within this mode is $(p_{k,i}^x,p_{k,i}^y,p_{k,i}^z)^T = \dot{A}_k(t)\sqrt{m_i}(c_{k,i}^x,c_{k,i}^y,c_{k,i}^z)^T.$ The sum over all the atoms of the product $p_{k,i}^zL_{k,i}^z$ thus yields the axial \textit{momentum pseudoscalar} (MPS) for mode $k$, defined as \begin{equation}\tag{4}\begin{aligned}\sum_i p_{k,i}^zL_{k,i}^z&=\sum_i[\dot{A}_k(t)\sqrt{m_i}c_{k,i}^z][\dot{A}_k(t) \sqrt{m_i}(x_i c^y_{k,i} - y_i c^x_{k,i})]\\
&=\dot{A}_k(t)^2\sum_im_ic_{k,i}^z(x_i c^y_{k,i} - y_i c^x_{k,i}).
\end{aligned}\label{PzLz}
\end{equation}
An obvious advantage of this expression as a measure of mode chirality is that while the $p^z_{k,i}$ and $L^z_{k,i}$ each average to zero over the normal mode's temporal period \footnote{With the exception of particular linear combinations of degenerate modes that can carry non-zero angular momentum \cite{cphonons_2D,cphonons_3D,phon_ang_mom}.} because they are first order in $\dot{A}_k(t)$, the product $p^z_{k,i}L^z_{k,i}$ is second order in $\dot{A}_k(t)$ and therefore has a well-defined sign.

Examining the density of modes with respect to the $\Sigma_i p_{k,i}^zL_{k,i}^z$ instead of the CCM yields profound results. First, \hyperref[FIG2]{Fig. 2(b)} shows that like the CCM, twist tends to skew the distribution of the axial MPS values away from zero, again indicating that chiral modes are associated with chiral structures. However, unlike the CCM, the sign of the MPS allows us to define handedness. This is illustrated in Fig. S8, which shows that while the CCM is even under spatial inversion (interchanging enantiomers), the MPS is odd under this transformation.

Secondly and remarkably, \hyperref[FIG2]{Fig. 2(b)} (which displays the mode density with respect to the momentum pseudoscalar) and \hyperref[FIG2]{Fig. 2(c)} (in which the momentum pseudoscalar averaged
over frequency bin is plotted against the mode frequency) indicate that different bands of the
frequency spectrum (\hyperref[FIG2]{Fig. 2(a)}) show different trends in developing handedness when the structure is twisted.

It is interesting to note that the form of the MPS (Eq. (\ref{PzLz})) is similar to that taken by the amplitude for
circular dichroism, $R(1_k \leftarrow 0)=\Sigma_\alpha\bra{0}\mu^\alpha\ket{1_k}\bra{0}\text{m}^\alpha\ket{1_k}$ (with $\mu^\alpha$ and $m^\alpha$ denoting the $\alpha$th Cartesian component of the electric and magnetic dipole operators respectively), in a
model where fixed partial charges are placed on each atomic cite \cite{Barron_reference,schellman_vibrational}. However, this case differs from the MPS by involving cross terms $(i\neq j)$.

Finally, we consider the equilibrium thermal average of the MPS, Eq. (\ref{PzLz}). Noting that $\dot{A}_k$ is the mass-weighted velocity coordinate of a harmonic oscillator, it satisfies $(1/2)\braket{\dot{A}_k^2}_T = (1/2)\hbar\omega_k\left(\braket{n(\omega_k)}_T+(1/2)\right)$ in quantum mechanics, and
$(1/2)\braket{\dot{A}_k^2}_T = (1/2)k_BT$
in the classical limit. Using this in Eq. (\ref{PzLz}) and summing over all modes $k$ defines a global quantity that we call the \textit{thermal chirality} ($\xi_z$), also a pseudoscalar. It is given by
\begin{equation}\tag{5a}\label{therm_chir_q}
\begin{aligned}
&\xi_z(T) = \sum_k \sum_i \braket{p^z_{k,i} L^z_{k,i}}_T\\
&=\sum_k \hbar\omega_k\left(\frac{1}{e^{\hbar\omega_k/k_BT}-1}+\frac{1}{2}\right) \sum_i m_ic^z_{k,i}(x_ic^y_{k,i}-y_ic^x_{k,i}),
\end{aligned}
\end{equation} which in the classical limit becomes
\begin{equation}\tag{5b}\label{therm_chir_c}
\begin{aligned}
&\xi_z(T)=k_BT \sum_k \sum_i m_ic^z_{k,i}(x_ic^y_{k,i}-y_ic^x_{k,i})=0.
\end{aligned}
\end{equation}
\begin{figure}[!hbt]
\includegraphics[width=1\linewidth]{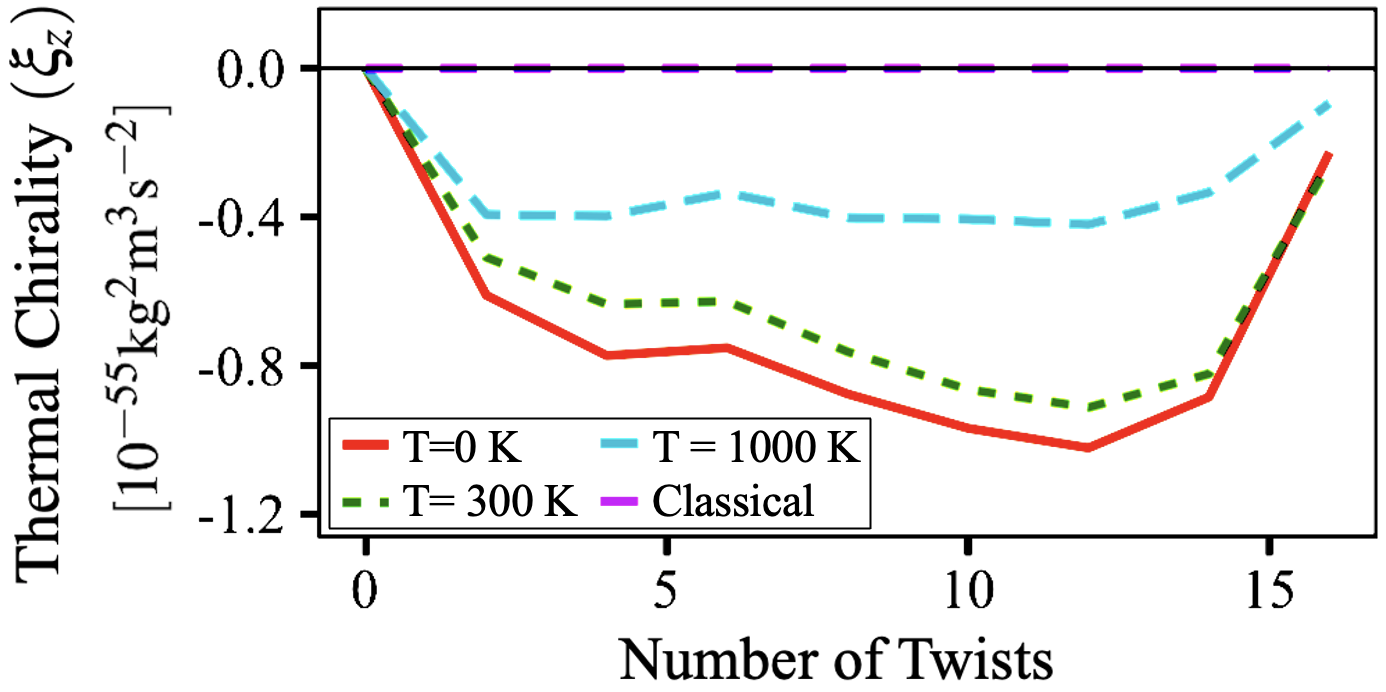}
\caption{Thermal chirality plotted as a function of the number of (left-hended) twists for the two-stranded $N=98$ polyethylene wire. Results are shown for (red) $T = 0$ K, (green) $T = 300$ K, (blue) $T = 1000$ K and (purple) classical limit.}
\label{FIG3}
\end{figure}The equality of Eq. (\ref{therm_chir_c}) to zero is due to the orthonormality of the $\{\textbf{c}_k\}$, which implies not only that $\sum_i\sum_\alpha c^\alpha_{k,i} c^\alpha_{k',i} = \delta_{k,k'}$, but also that $\sum_k c^\alpha_{k,i} c^\beta_{k,j} = \delta_{i,j}\delta_{\alpha,\beta}$, where $\delta_{m,n}$ is the Kronecker delta function. From this, we suggest that thermal chirality is a strictly quantum phenomenon, given by Eq. (\ref{therm_chir_q}).

\hyperref[FIG3]{Figure 3} shows this thermal chirality measure as a function of the number of twists. We see that it is more significant in the low temperature limit, as expected for a quantum phenomenon. The fact that the thermal chirality is nonzero for twisted structures is an example of the breakdown of the equipartition theorem in the quantum regime, for it suggests a nonzero cross-correlation of degrees of freedom $\langle \dot{u}_i^x \dot{u}_i^z\rangle \neq 0$ and $\langle \dot{u}_i^y \dot{u}_i^z\rangle \neq 0$ at equilibrium.

In conclusion, we have introduced measures for the chirality of molecular normal modes and examined their spectral properties and their dependence on the chirality of the underlying molecular structure. We note that qualitative correlations between chirality measures of structures (mainly the CCM) and molecular chiral responses have already been demonstrated \cite{trans_metal,chral_nanoalloy,chir_trans_enthalpy,chir_enantio}, while correlation between atomic linear and angular momenta has already been discussed as a source of anisotropy in phonon transport in crystalline chiral solids \cite{chir_phon_materials,cphonons_3D}. The correlation between the linear and angular momentum in the modes of the double helix demonstrated in this work suggests the possibility that a particle interacting with the system could exhibit correlations between its exchange of linear and angular momentum with the system, providing a classical model for chiral friction. This as well as further implications of the MPS to molecular VCD, and the possibility that thermal chirality (Eq. (\ref{therm_chir_q})) may be related to observations of emission of circularly polarized radiation \cite{incandescent_CPL,blackbody_CPL} or excitation of chiral phonons by thermal gradients \cite{therm_angular} will be the subject of future studies.

{\it Acknowledgements.} The research of A.N. is supported by the Air Force Office of Scientific Research under award number FA9550-23-1-0368 and the University of Pennsylvania. E.A. aknowledges the support of the the University of Pennsylvania (Grant GfFMUR). The authors are grateful to Mohammadhasan Dinpajooh and Claudia Climent for technical assistance, and to Pere Alemany, David Avnir, Adam Cohen, Oded Hod, Randall Kamien, and Philip Nelson for useful discussions.

{\it Data Availability.} Normal modes for sample structures as in \hyperref[FIG1]{Fig. 1} are available on github at: 
\url{https://github.com/eabes23/chiral_modes/}. The codes used to generate the twisted structures are also available at: \url{https://github.com/eabes23/polymer_twist/}. Codes for applying the CCM are open source at: \url{https://zenodo.org/records/4925767} and at: \url{https://cosymlib.readthedocs.io/en/pere_tutorial/}. Further data and analysis codes are available upon reasonable request. See the Supplemental Material \footnote{See Supplemental Material [url], which
includes Refs. \cite{twist_paper,Hadi,FJC, FF3,FF5,unitedWorks, phon_ang_mom, participationRatio2}, for methods and further analysis.} for methods.

\bibliographystyle{aipnum4-1}

\clearpage
%%%%%%%%%% Merge with supplemental materials %%%%%%%%%%
\pagebreak
\widetext
\begin{center}
\textbf{\large Supplemental Information:\\ Quantifying the Chirality of Vibrational Modes in Helical Molecular Chains}
\end{center}
%%%%%%%%%% Merge with supplemental materials %%%%%%%%%%
%%%%%%%%%% Prefix a "S" to all equations, figures, tables and reset the counter %%%%%%%%%%
\setcounter{equation}{0}
\setcounter{figure}{0}
\setcounter{table}{0}
\setcounter{section}{0}
\setcounter{page}{1}
\makeatletter
\renewcommand{\theequation}{S\arabic{equation}}
\renewcommand{\thefigure}{S\arabic{figure}}
\renewcommand{\bibnumfmt}[1]{[S#1]}
\renewcommand{\citenumfont}[1]{S#1}
\renewcommand{\thepage}{S\arabic{page}} 
%%%%%%%%%% Prefix a "S" to all equations, figures, tables and reset the counter %%%%%%%%%%

\section{Computational Details}
\label{sec1}

A detailed description of how the twisted structures and normal modes were obtained can be found in Ref. \cite{twist_paper}. The twisted structures were obtained from MD simulations in which a torque was applied to one end of the polymer wire while keeping the other end fixed using LAMMPS. Normal modes and corresponding eigenfrequencies were calculated from energy minimized structures using GROMACS. Normal modes associated with imaginary frequencies were discarded from the analyses, but for all structures examined these composed $<6\%$ of the spectrum. The CCM calculations were implemented using the \texttt{gsym} function of the \texttt{cosymlab} library in Python. All CCM calculations assumed that the nearest achiral structure had $S_1$ symmetry, and the trivial permutation was forced in order to respect the information contained in bond connectivity. Calculations were repeated on multiple configurations sampled from the MD simulations and thermal fluctuations were found to be minor. The polymers were studied at their natural untwisted length by fixing terminal atoms at either end. The force field used to model the polymers in the main text was the TraPPE-UA force field, but conclusions were cross-checked using other force fields as shown in the supplementary material.

\section{Molecular Models}
\label{sec2}

For the calculations in the main text, we have representing our polymers using the Transferable Potentials for Phase Equilibria (TraPPE) United Atom (UA) model, which treats each CH$_x$ unit as a single particle \cite{Hadi, FF3, FF5}. This model is useful because it greatly reduces the computational cost, yet it has been shown to still perform with high accuracy for calculations of thermal conductance in hydrocarbons \cite{unitedWorks}. It is based on a force field (FF) that represents bonds with harmonic potentials and includes also angle (3-body), dihedral (4-body potentials), and Lennard Jones potentials. The Hamiltonian is given by\begin{equation}
\begin{split}
H_{\rm{molecule}} & = \sum_i \frac{p_i^2}{2m_i} +  \sum_i k_{bi} (l_i-l_{0i})^2 + \sum_i k_{\theta i} (\theta_i-\theta_{0i})^2 + \\
  & \sum_i \sum_{n_i}^4 \frac{C_{n i}}{2} \left[ 1+ (-1)^{n{_i}-1} {\rm{cos}} ( n_i \phi_i) \right] + \sum_i \sum_j 4 \epsilon_{ij} \left[  (\frac{\sigma_{ij}}{r_{ij}})^{12} - (\frac{\sigma_{ij}}{r_{ij}})^{6}  \right],
\end{split}
\label{Hmol}
\end{equation}where $p_i$ and $m_i$ are the momentum and mass of a given particle respectively, $l_i$,$l_{0i}$, $\theta_i$,$\theta_{0i}$, are the actual and equilibrium bond lengths and angles respectively, $\phi_i$ are the dihedral angles, and $r_{ij}$ are the inter-particle separations. Accordingly, the parameters $k_{bi}$ and $k_{\theta i}$ are the spring constants for the bonds and angles, and the $C_{ni}$ are the constants that define the dihedral potentials. As in Ref. \cite{Hadi}, the above parameter values were chosen to fit observed physical properties.

As we have done in our previous work \cite{twist_paper}, we have cross-checked our results with two other models: i) an Explicit Hydrogen (EH) model and ii) a simplified force field inspired by the Freely Joint Chain (FJC) model \cite{FJC}. The Explicit Hydrogen model uses the same Hamiltonian as above, except that the Hydrogen and Carbon atoms are treated as separate bodies and different parameter values are chosen as appropriate (see  Appendix A in \cite{twist_paper}). The FJC force field is also a united atom model that conventionally omits the angular potentials, dihedral potentials, and Leonard Jones (LJ) potentials. In the present application, since our calculations pertain to wires with multiple strands, the omission of the Leonard Jones potentials would create a non-physical situation with no repulsive force between two chains. We, therefore, modify the FJC field to include the LJ potential and denote this force field as FJC$^*$. Hence the Hamiltonian becomes \begin{equation}
H_{\rm{FJC^*}} = \sum_i \frac{p_i^2}{2m_i} +  \sum_i k_{bi} (l_i-l_{0i})^2 + \sum_i \sum_j 4 \epsilon_{ij} \left[  (\frac{\sigma_{ij}}{r_{ij}})^{12} - (\frac{\sigma_{ij}}{r_{ij}})^{6}  \right],
\label{FRC}
\end{equation}using the same parameter values as the TraPPE-UA FF. It should be noted that this FJC$^*$ model allows the angles to relax from $\theta_0=114^{\circ}$ to $180^{\circ}$, changing the natural length of the polymer. In our previous work we reported FJC$^*$ results using multiple lengths \cite{twist_paper}, but here for all FJC$^*$ results reported, we have adjusted the length to the new natural length of the polymer, which is equal to the contour length of the polymer when using the TraPPE-UA model.

\newpage
\pagebreak
\newpage	

\section{CCM Normal Mode Spectra}
\label{sec3}

In our main text we presented the key finding that increased twist skews the distribution of the CCM of normal modes away from zero to greater mean values. Here we show that this trend was remarkably consistent across various trials, chain lengths, and force field models. \begin{figure*}[!hbt]
\includegraphics[width=0.8\linewidth]{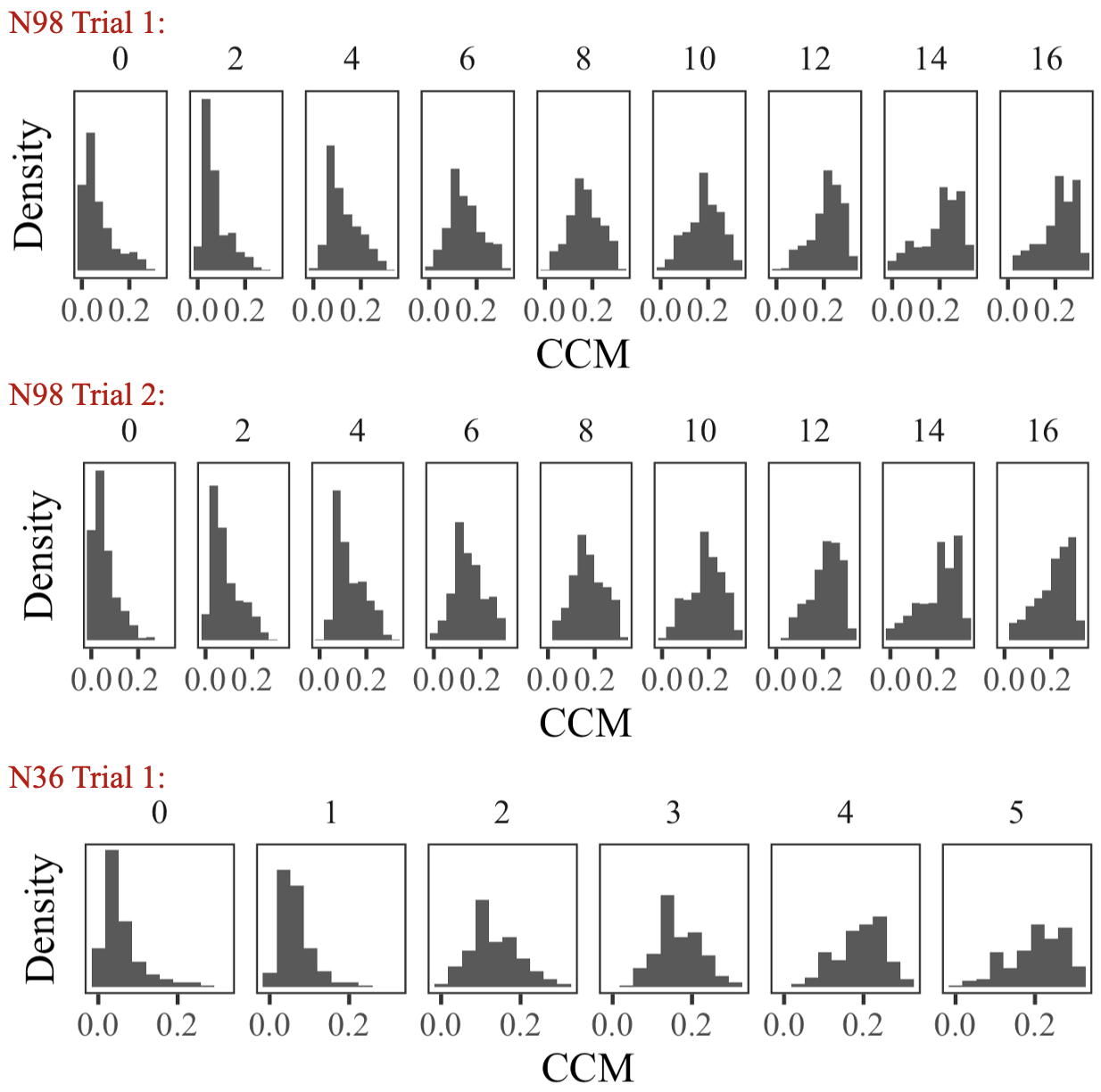}
\caption{(Top and middle) Histograms of the normal mode distribution binned by CCM for two stranded polyethelene wires of length $N=98$ for various levels of twist. The number above each plot denotes the number of twists. Results are shown using two configurations of the same molecular structure sampled from two arbitrary timepoints throughout MD simulations. (Bottom) Same as above but for one trial of a polymers of length $N=36.$}
\label{FIG1}

\end{figure*}The top and middle panels of \hyperref[FIG1]{Fig. S1} show that the trend is consistent across trials. Here, different \textit{trials} denote different configurations lifted from identically prepared molecular dynamics (MD) simulations. For a detailed description of such MD simulations, see our prior work \cite{twist_paper}. Although normal modes were computed from energy minimized structures, for molecules of this size there exist multiple local minima that have slightly different normal modes. We see that although the relative size of individual histogram bars varies slightly, there is no observable difference in the trend we have reported. Ten such trials have been checked (results not shown).
The bottom panel shows that the same trend persists for a shorter chain length ($N=36$). 

 \begin{figure*}[!htb]
\includegraphics[width=0.5\linewidth]{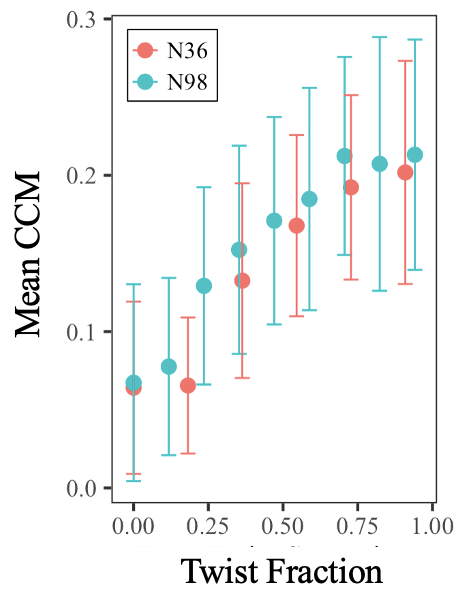}
\caption{The mean of the normal mode CCM distributions shown in trial 1 of \hyperref[FIG1]{Fig. S1} as a function of the Twist Fraction given by $N_T/N_T^{\text{max}}(N)$. Results are shown for lengths (red) $N=36$ and (blue) $N=98.$ Error bars show the standard deviation of the distributions.}
\label{FIG3}
\end{figure*}

Note that as one would expect, we cannot obtain as great an absolute number of twists $N_T$ in the shorter chain as we could with a larger chain length. Such wires have been found to be characterized by a maximal number of twists $N^{\max}_T(N)$ before bonds begin to break. As shown in our previous work \cite{twist_paper}, this is quantity $N^{\max}_T(N)$ is proportional (at least to a very good approximation) to the chain length of $N$. As such, when assessing the effect of twist on physical properties, the twist fraction $N_T/N^{\max}_T(N)$ is likely more fundamental than the absolute number of twists. \hyperref[FIG1]{Figure S2} shows that with respect to this parameter, the effect of twist on the mean of the distributions is consistent across chain lengths. Note that the CCM of these modes therefore appears to be independent of the absolute number of atoms in the structure. This was not the case when the CCM was applied to the \textit{structure} as opposed to the modes, as discussed in the main text.

 \begin{figure*}[!htb]
\includegraphics[width=0.8\linewidth]{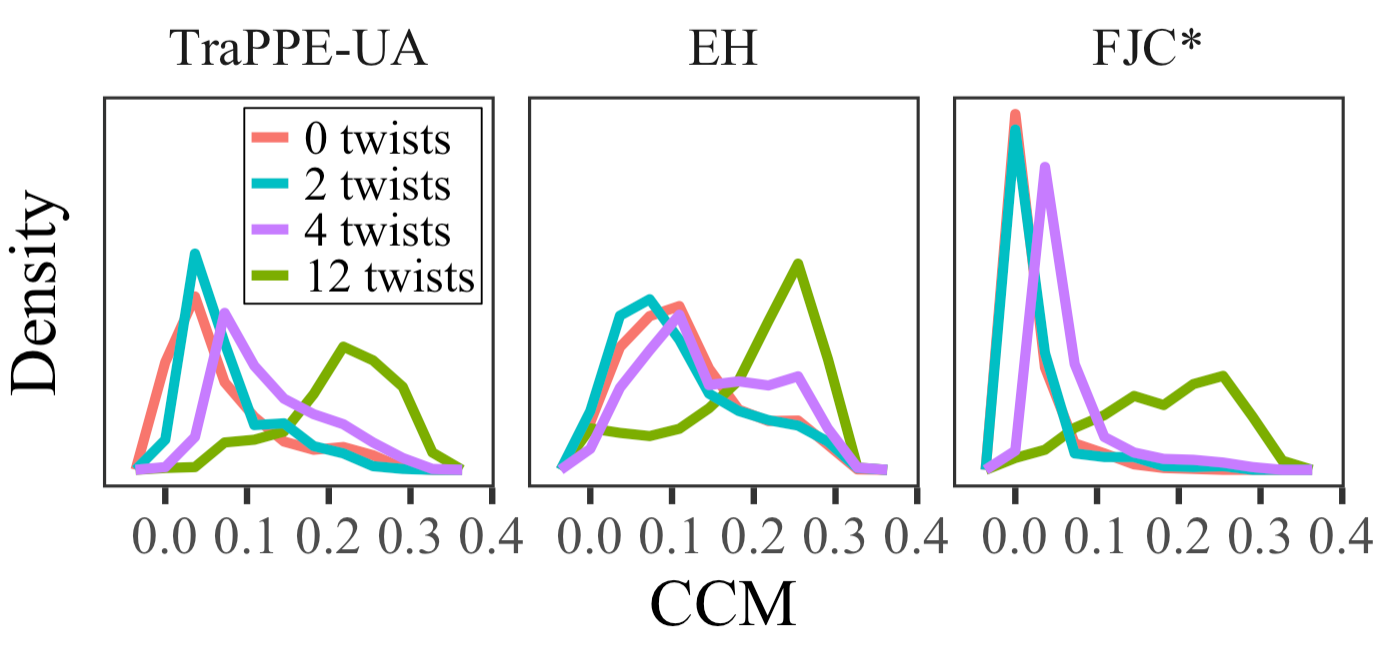}
\caption{Frequency polygons of normal modes binned by CCM for two-stranded polyethelene wires of length $N=98$ at various levels of twist. Results are shown for (red) untwisted wire, (blue) 2 twists, (purple) 4 twists, and (green) 12 twists. Results are compared for (left) the TraPPE-UA model used in the main text, (center) an Explicit Hydrogen model, and (right) a variation of the Freely Joint Chain model (see \hyperref[sec2]{Sec. II} for details).}
\label{FIG3}
\end{figure*}

In addition to checking various lengths, we checked whether the trend persists across various force fields models. This is the subject of \hyperref[FIG3]{Fig. S3} which compares (left) the results of the main text to analogous results using the other examined force fields (see \hyperref[sec2]{Sec. II}). The center panel shows an Explicity Hydrogen model which does not make the United Atom simplification. We see that the same trend, the shifting of the CCM distribution with twist, persists with this model as well. The right panel shows that the trend also persists using the FJC$^*$ model, which is the most reductionistic of the three examined.

 \begin{figure*}
\includegraphics[width=0.8\linewidth]{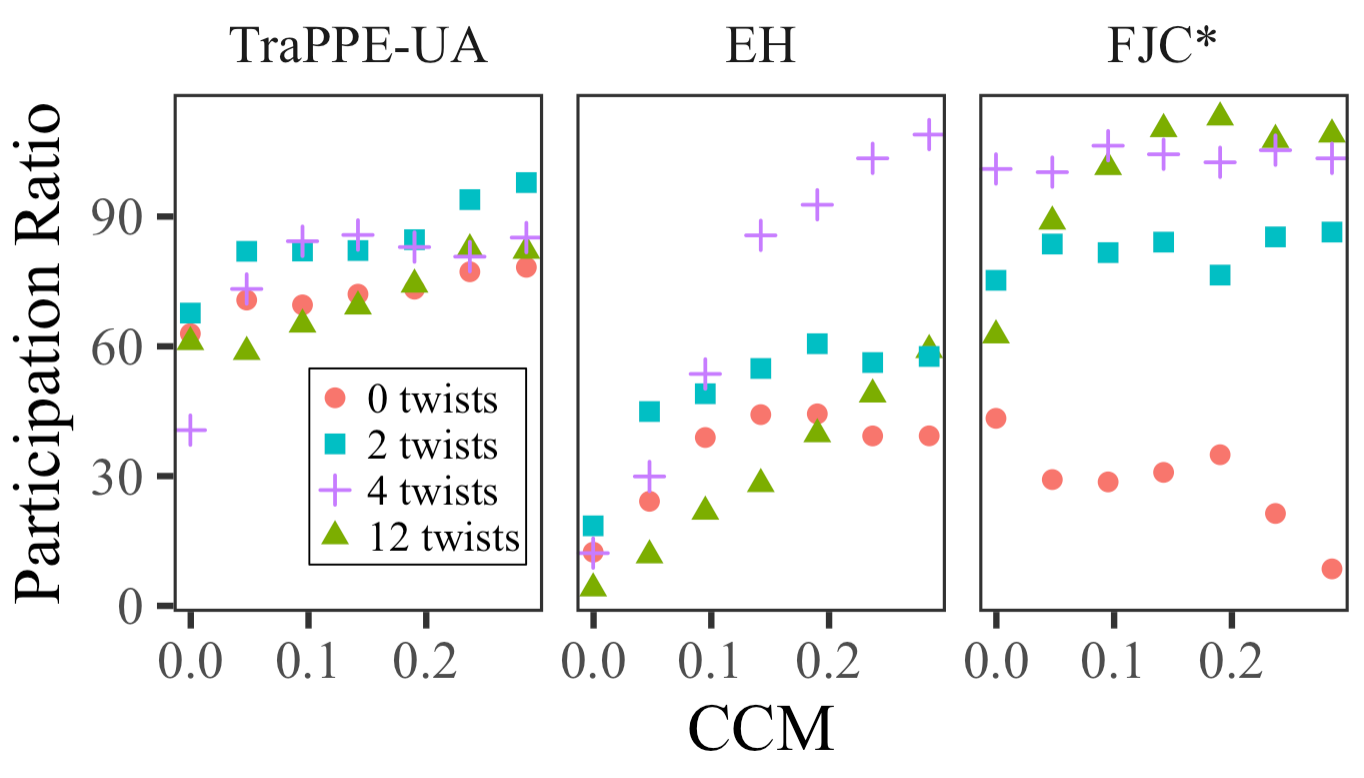}
\caption{Participation ratio coarse-grained by CCM bin for Frequency polygons of the normal mode distributions binned by CCM for two-stranded polyethelene wires of length $N=98$ at various levels of twist. Results are shown for (red circles) untwisted wire, (blue squares) 2 twists, (purple crosses) 4 twists, and (green triangles) 12 twists. Results are compared for (left) the TraPPE-UA model used in the main text, (center) an Explicit Hydrogen model, and (right) a variation of the Freely Joint Chain model.}
\label{FIG4}
\end{figure*}

 We have also found a correlation between CCM and mode localization. A canonical localization measure for normal modes is the \textit{participation ratio ($P_k$)} of mode $k$, varying from $1$ when the vibrations are localized on a single atom to $N$ if the mode is equally distributed on all atoms. The participation ratio is computed as follows: let the coefficients $c^\alpha_{k,i}$ denote the expansion of normal mode $k$ in the atomic coordinates where $\alpha$ denotes the Cartesian coordinate and $i$ denotes the atom number. Defining $p_{k,i}=\sum_{\alpha} |c_{k,i}^\alpha|^2$, the participation ratio is then defined as \begin{equation}\tag{4}\label{eq:Pk}P_k = 1/\sum^N_i p^2_{k,i}.\end{equation} Note that since we have taken the modes to be normalized, i.e. $\sum_i \sum_\alpha |c_{k,i_\alpha}|^2 = \sum^N_i p_{k,i}=1$ \cite{twist_paper, participationRatio2}.

\hyperref[FIG4]{Figure S4} shows that across a variety of force fields, the participation ratio is strongly correlated with the CCM in $N=98$. This trend is persistent across various levels of twist, except for the untwisted wire in the FJC$^*$ model (note that the FJC$^*$ model is the least detailed and thus least physically accurate of the three force fields examined). This result suggests that the chiral modes could be invloved in global transport properties of the molecule.

\newpage
\pagebreak
\newpage

\section{Phonon Angular Momentum}
\label{sec5}

We follow Ref. \cite{phon_ang_mom} in deriving an expression for the phonon-angular momentum, defined by the atomic motion relative the the equilibrium coordinates rather than the central axis. The general form for the phonon angular momentum is \begin{equation}
\textbf{J} = \sum_i \textbf{u}_i \times \dot{\textbf{u}}_i
\label{angular}
\end{equation}
where $\textbf{u}_i$ and $\dot{\textbf{u}}_i$ are the mass-weighted displacement from equilibrium and velocity of atom $i$. We consider the $z$-component $J^z=\sum_i (u_i^x\dot{u}_i^y - u_i^y\dot{u}_i^x),$ and we use second quantization to write $u_i^\alpha$ as
\begin{equation}\tag{S4a}
u_i^\alpha = \sum_k c_{k,i}^\alpha e^{-i\omega_k t}\sqrt{\frac{\hbar}{2\omega_k}}a_k+\text{h.c.}
\label{u}
\end{equation}
\begin{equation}\tag{S4b}
\dot{u}_i^\alpha = \sum_k -ic_{k,i}^\alpha e^{-i\omega_k t}\sqrt{\frac{\hbar\omega_k}{2}}a_k+\text{h.c.,}
\label{udot}
\end{equation}
where $c_{k,i}^\alpha$ is the mode coefficient described in the main text and $a_k$ is the annihilation operator of the harmonic mode $k$ (note that $a_k$ is related to $A_k$ from the main text by $a_k = \sqrt{\frac{\omega_k}{\hbar}}A_k)$.

Substituting Eq. (\ref{u}) and Eq. (\ref{udot}) into the expression for $J_i^z$, we obtain 
\begin{equation}\tag{S5}\begin{split} 
J_i^z = \sum_k\sum_{k'} \left[ \left(c_{k,i}^x e^{-i\omega_k t}\sqrt{\frac{\hbar}{2\omega_k}}a_k+\text{h.c.}\right)\left(-ic_{k',i}^y e^{-i\omega_{k'} t}\sqrt{\frac{\hbar\omega_{k'}}{2}}a_{k'}+\text{h.c.}\right)\right.\\-\left.\left(c_{k,i}^y e^{-i\omega_k t}\sqrt{\frac{\hbar}{2\omega_k}}a_k+\text{h.c.}\right)\left(-ic_{k',i}^x e^{-i\omega_{k'} t}\sqrt{\frac{\hbar\omega_{k'}}{2}}a_{k'}+\text{h.c.}\right)\right]
\label{Jz}
\end{split}\end{equation}
The $a_k a_k$ and $a_k^\dagger a_k^\dagger$ do not contribute at equilibrium due to the fast oscillations. Therefore, we have
\begin{equation}\tag{S6}\begin{split} 
J_i^z = \sum_k\sum_{k'} \frac{i\hbar}{2} \sqrt{\frac{\omega_{k'}}{\omega_k}}\left[c_{k,i}^x c_{k',i}^{y*}e^{-i(\omega_k-\omega_{k'}) t}a_k a_{k'}^\dagger -c_{k,i}^{x*} c_{k',i}^ye^{-i(\omega_{k'}-\omega_k) t}a_k^\dagger a_{k'}\right. \\ -\left. c_{k,i}^y c_{k',i}^{x*}e^{-i(\omega_k-\omega_{k'}) t}a_k a_{k'}^\dagger +c_{k,i}^{y*} c_{k',i}^xe^{-i(\omega_{k'}-\omega_k) t}a_k^\dagger a_{k'}
\right],
\label{Jz}
\end{split}\end{equation} which by appropriate rearranging and switching of dummy indices becomes 
\begin{equation}\tag{S7}\begin{split}
J_i^z = \sum_k\sum_{k'}\frac{i\hbar}{2}\left[(c_{k,i}^x c_{k',i}^{y*} -  c_{k,i}^y c_{k',i}^{x*})\sqrt{\frac{\omega_{k'}}{\omega_k}}e^{-i(\omega_k-\omega_{k'})t}a_k a_{k'}^\dagger\right. \\ - (c_{k',i}^{x*} c_{k,i}^y -  c_{k',i}^{y*} c_{k,i}^x)\left.\sqrt{\frac{\omega_{k}} {\omega_k'}}e^{-i(\omega_k-\omega_{k'})t}a_{k'}^\dagger a_k \right]
\end{split}\label{Jz2}\end{equation}

$$=\sum_k\sum_{k'}\frac{i\hbar}{2}(c_{k,i}^x c_{k',i}^{y*} -  c_{k,i}^y c_{k',i}^{x*})\left(\sqrt{\frac{\omega_{k'}}{\omega_k}}a_k a_{k'}^\dagger+\sqrt{\frac{\omega_{k}} {\omega_k'}}a_{k'}^\dagger a_k\right)e^{-i(\omega_k-\omega_{k'})t}. $$

At this point, we can use $[a_k,a_{k'}^\dagger]=\delta_{k,k'}$ and sum over all atoms $i$ to obtain

\begin{equation}\tag{S8}
J^z = \sum_{k}\sum_{k'}\sum_{i}\frac{i\hbar}{2}(c_{k,i}^x c_{k',i}^{y*} -  c_{k,i}^y c_{k',i}^{x*})\left(\sqrt{\frac{\omega_{k'}}{\omega_k}}+\sqrt{\frac{\omega_{k}} {\omega_k'}}\right)e^{-i(\omega_k-\omega_{k'})t}a_{k'}^\dagger a_k.
\label{Jz_3}\end{equation}

Our last step is to consider the thermal average $\langle{J^z}\rangle_T,$ which depends on $\langle{e^{-i(\omega_k-\omega_{k'})t}a_{k'}^\dagger a_k}\rangle_T=\delta_{k,k'}\langle{n({\omega_k})}\rangle_T,$
%\begin{equation}\tag{S9}\langle{e^{-i(\omega_k-\omega_{k'})t}a_{k'}^\dagger a_k}\rangle_T=\[ \begin{cases} 
      %0, & |\omega_k - \omega_{k'}| > 0 \\
      %\langle{n({\omega_k})}\rangle_T,
      %& \omega_k \approx \omega_{k'},
   %\end{cases}
%\]
%\label{peicewize}\end{equation}
where $\langle{n({\omega_k})}\rangle_T$ is the frequency-dependent thermal average occupancy. We find

\begin{equation}\tag{S9}
J^z = 2\hbar \sum_{k}\langle{n({\omega_k})}\rangle_T\sum_{i}\text{Im}[c_{k,i}^{x*} c_{k,i}^y],
\label{Jz_final}\end{equation}
which we note is equal to zero if the $c_{k,i}^{\alpha}$ are real.

\newpage
\pagebreak
\newpage	

\section{Momentum Pseudoscalar Spectra}
\label{sec5}

\begin{figure*}[!htb]
\includegraphics[width=0.8\linewidth]{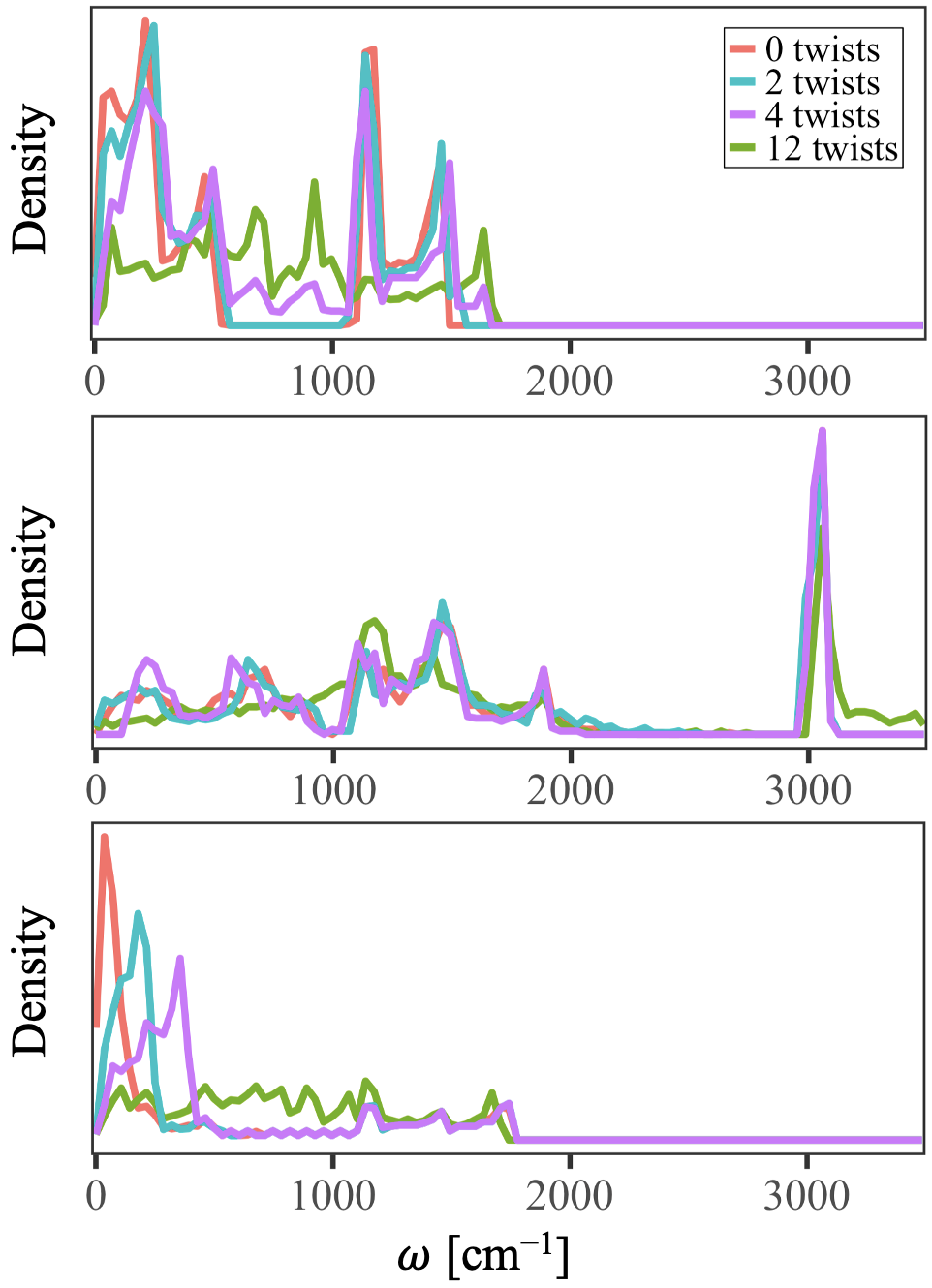}
\caption{Spectral densities binned by frequency for two-stranded polyethelene wires of length $N=98$ at various levels of twist. Results are shown for (red) untwisted wire, (blue) 2 twists, (purple) 4 twists, and (green) 12 twists. Results are compared for (top) the TraPPE-UA model used in the main text, (middle) an Explicit Hydrogen model, and (bottom) a variation of the Freely Joint Chain model (see \hyperref[sec2]{Sec. II} for details).}
\label{FIG5}
\end{figure*}

In this section we demonstrate the persistence across various chain lengths and force fields of the second key finding of our main text, that momentum pseudoscalar (MPS) score distribution is shifted in different directions for different frequency bands. \hyperref[FIG5]{Figure S5} shows the spectral densities as a function of frequency for various levels of twist using three different force fields. The top panel shows data from the main text. Note the band gap (roughly between $\omega=6$ cm$^{-1}$ and $\omega=1100$ cm$^{-1}$) for 0 twists and 2 twists (red and blue line respectively). This band gap begins to disappear for 4 twists (purple line) and fully disappears for 12 twists (green line). For both other force fields, a similar band gap that fades with increased twist is present albeit less obvious. For the Explicit Hydrogen model (middle) this band gap is much narrower (appears at roughly $\omega=1000$ cm$^{-1}$), and for the FJC$^*$ model (bottom), the density of the high frequency band is greatly diminished. Note that in a structure with $n$ particles and fixed center of mass, there are $3N$ normal modes (as mentioned in the main text, a small fraction returned imaginary frequencies; we discarded these from our analysis). Since there are $n=588$ atoms in an $N=98$ polyethylene double helix, this implies that there are 1764 total modes for the Explicit Hydrogen model. For the other force field models, which makes the United Atom simplification, there are only $n=196$ total atoms and thus 588 normal modes. Because these details distract from our main point which is the shape of the distributions, we here and throughout the entire work we label our vertical axis with \textit{density} (arb. u.). Note also that in only the EH model, there is a spike in the spectral density at roughly $\omega=3000$ cm$^{-1}.$ We attribute this to modes involving motions of hydrogen atoms with minimal participation from the carbon atoms. 

\begin{figure*}[!h]
\includegraphics[width=0.8\linewidth]{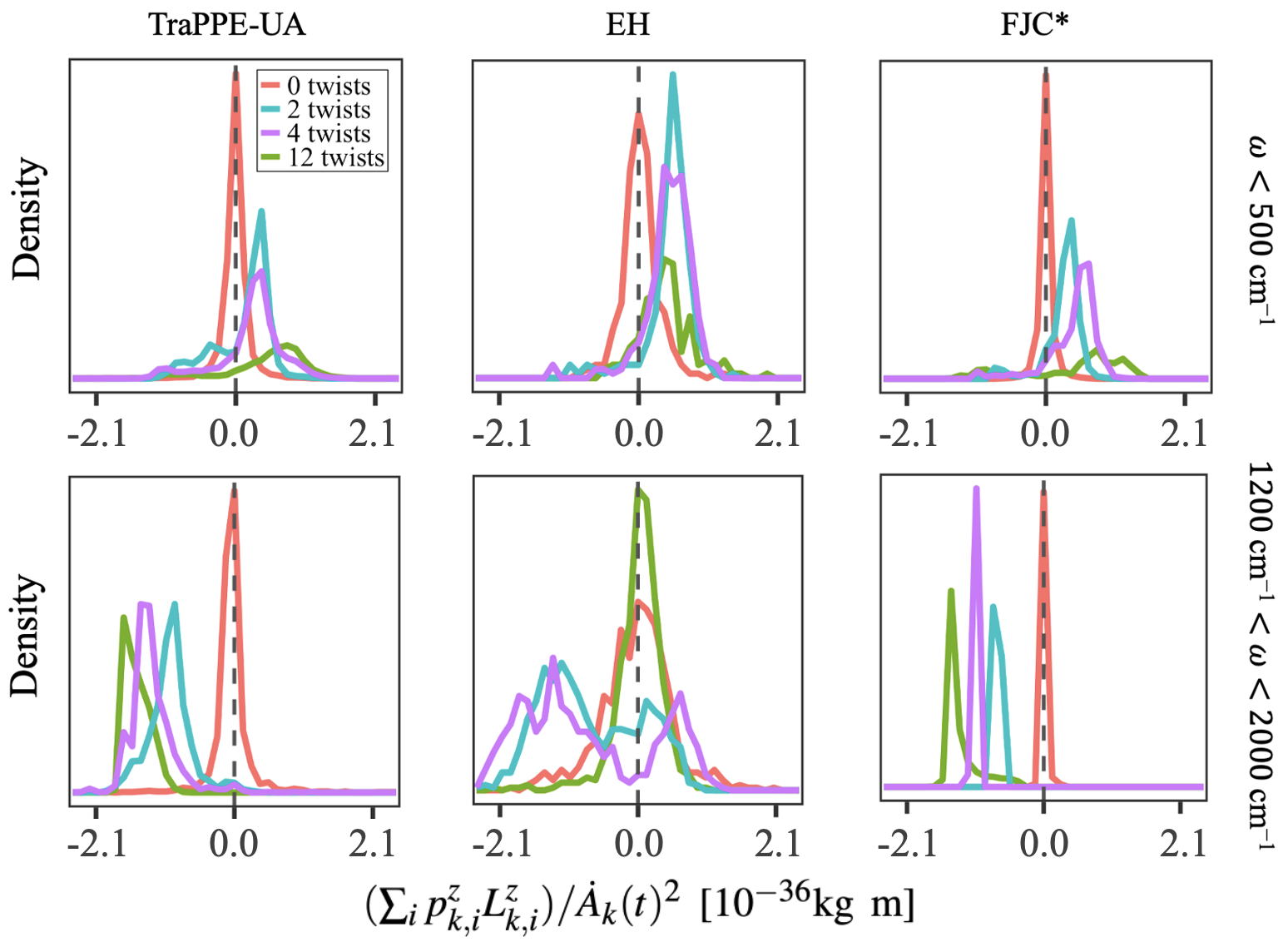}
\caption{Frequency polygons of normal modes binned by axial momentum pseudoscalar for $N=98$ polyethelene double helical wires over particular frequency ranges. The (top) low frequency range is $\omega<500$ cm$^{-1}$ and the (bottom) high frequency range is $1200$ cm$^{-1}<\omega<2000$ cm$^{-1}.$ Results are shown for (red) untwisted wire, (blue) 2 twists, (purple) 4 twists, and (green) 12 twists. Results are compared for (left) the TraPPE-UA model used in the main text, (center) an Explicit Hydrogen model, and (right) a variation of the Freely Joint Chain model .}
\label{FIG6}
\end{figure*}

\hyperref[FIG6]{Figure S6} Shows the normal mode densities with respect to the axial MPS (Eq. (4) in main text) for two two frequency ranges. The left panels were also shown in the main text. Remarkably, in all three models, when the wire is twisted, the peak of the low frequency bands shifts to the right and that of the high frequency band shifts to the left. Note that although the peaks for the more detailed EH model (center) are less clearly defined, the general trend persists. The one exception is that in the EH model it appears that for the highly twisted structure (12 twists), the low frequency distribution shifts back towards zero. Lastly, note that for the highest frequency band in the Explicit Hydrogen model $(\omega\sim3000$ cm$^{-1})$, the corresponding peaks shifts to the right (result not shown). In total, these results suggest that structural chirality gives handedness to bands of the material's frequency spectrum. This is the result which has lead us to introduce the concept of \textit{thermal chirality}.

\begin{figure*}[!h]
\includegraphics[width=0.9\linewidth]{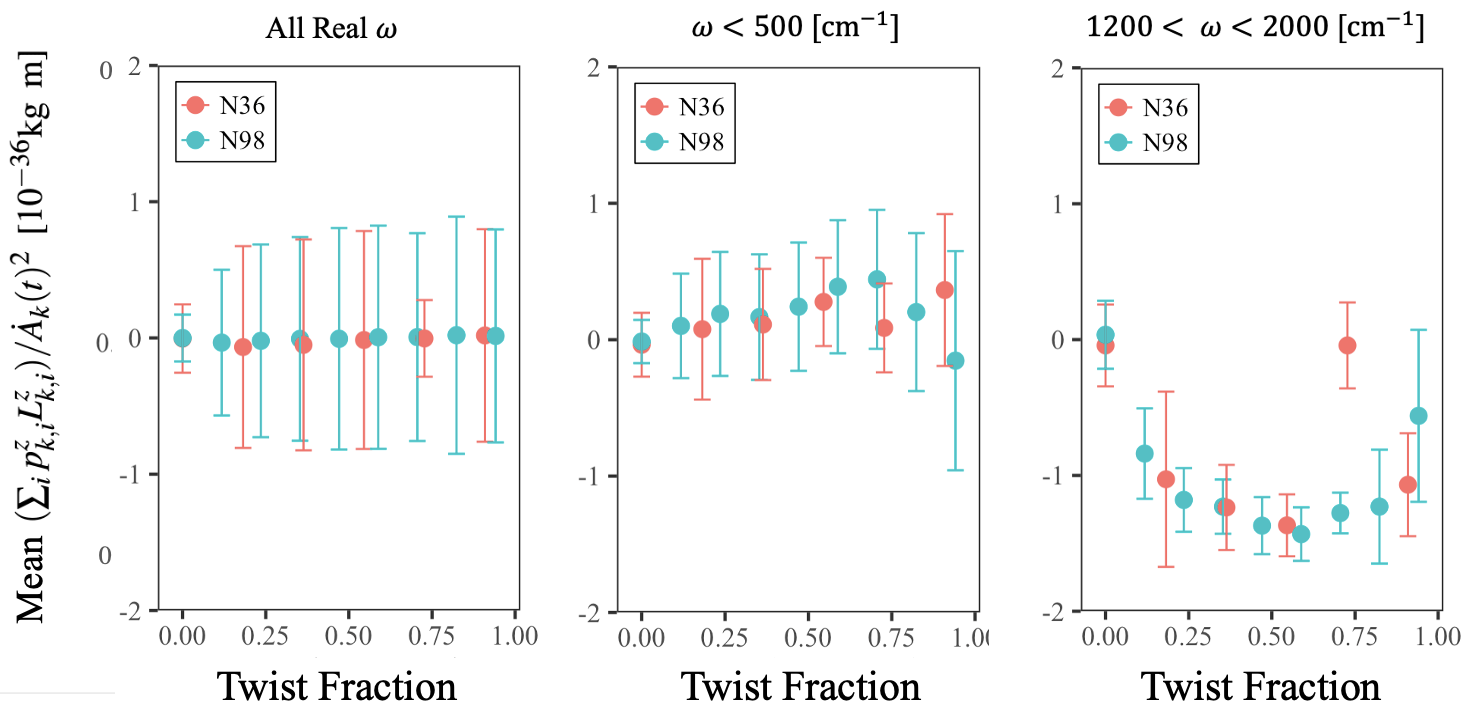}
\caption{The mean of the normal mode axial distributions for the TraPPE-UA force field as a function of the Twist Fraction given by $N_T/N_T^{\text{max}}(N)$. Results are shown for lengths (red) $N=36$ and (blue) $N=98,$ and over specific frequency ranges: (left) full spectrum, (center) $\omega<500$ cm$^{-1}$, and (right) $1200$ cm$^{-1}<\omega<2000$ cm$^{-1}.$ Error bars show the standard deviation of the distributions.}
\label{FIG7}
\end{figure*}

\hyperref[FIG7]{Figure S7} shows the mean and standard deviations of the distributions in these trends persist across various chain, and that the trends shown in \hyperref[FIG6]{Fig. S6} persist across different chain lengths. \hyperref[FIG7]{Figure S7} also shows that although individual frequency bands are shifted by twist (center and right), the mean of the full spectrum remains near zero (left). It is therefore often necessary to observe individual frequency bands in order to find the chiral characteristics of normal mode spectra. 

\clearpage

\section{Behavior of the Normal Mode Chirality Measures Under Parity}
\label{sec6}

From the definition of the CCM (Eq. (1) in the main text), it follows that the CCM is the same for both enantiomers, while the momentum pseudoscalar (Eq. (4) in the main text) is odd under conversion between enantiomers. Formally, this is equivalent to the statement that the CCM is even under spatial inversion ($\mathcal{P}$), while the momentum pseudoscalar $\Sigma_i p_{k,i}^zL_{k,i}^z$ is odd under this transformation.

\begin{figure*}[!h]
\includegraphics[width=0.5\linewidth]{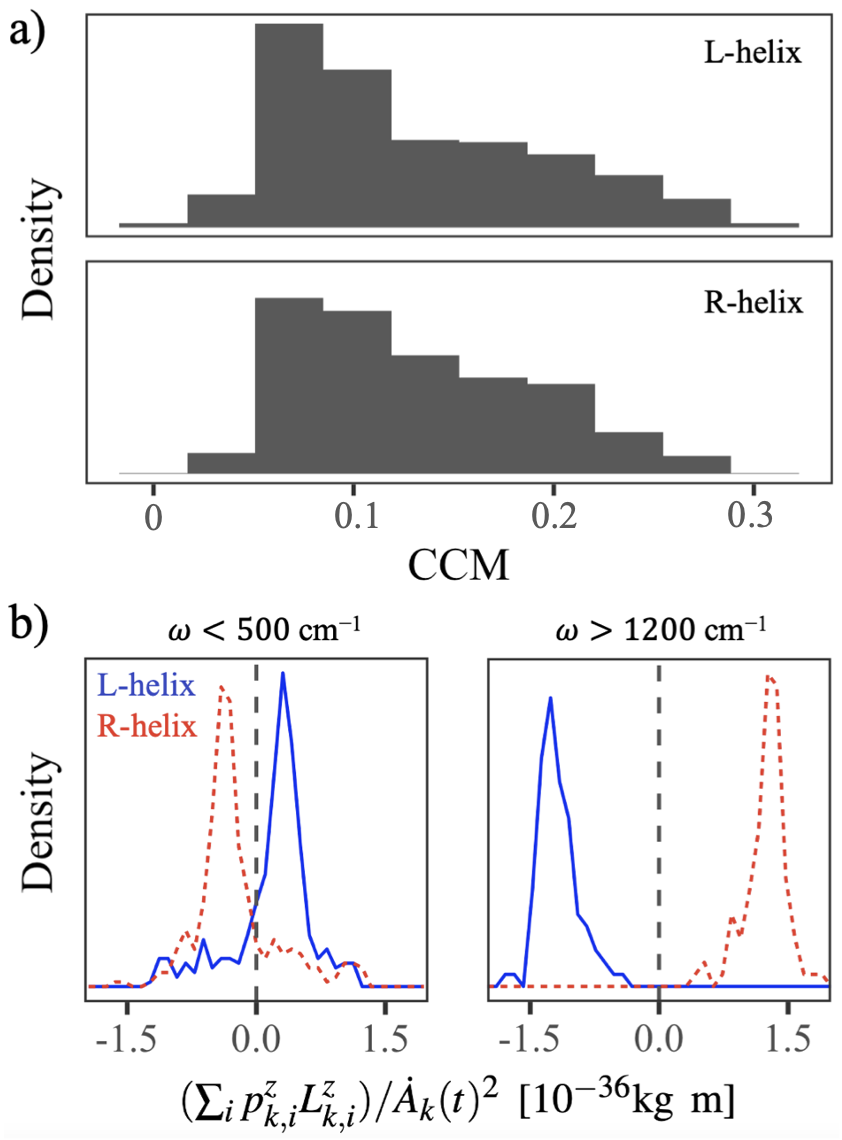}
\caption{(a) Comparison of the histograms of the normal mode distribution binned by CCM for (top) left-handed and (bottom) right-handed two-stranded polyethylene wires of length $N=98$ containing four twists. (b) Frequency polygons showing the density of modes binned by an axial momentum pseudoscalar score for the same structures as the top panel. The left panel plots only the modes from the low frequency band, and the right from the high frequency band. The solid blue line denotes the result for the left-handed helix shown in the main text, and the dotted red line denotes that for the corresponding right-handed helix.}
\label{FIG8}
\end{figure*}

While these formal relationships are transparent, it is reassuring to confirm them numerically. Here we perform identical calculations on both the left-handed and right-handed enantiomers from our study, containing four twists each. \hyperref[FIG8]{Figure S8(a)} confirms that the distribution of normal modes binned by CCM is qualitatively the same for either enantiomer, while \hyperref[FIG8]{Fig. S8(b)} conforms that the distribution of normal modes binned by momentum pseudoscalar is related by a negative sign between the enantiomers. Note that while the qualitative relationships between the distributions are as predicted, there is expectedly a small amount of noise due to thermal fluctuations in the MD simulations used to generate the structures.

\clearpage
\section{Inter-Chain Distance}
\label{sec7}

\begin{figure*}[!h]
\includegraphics[width=0.5\linewidth]{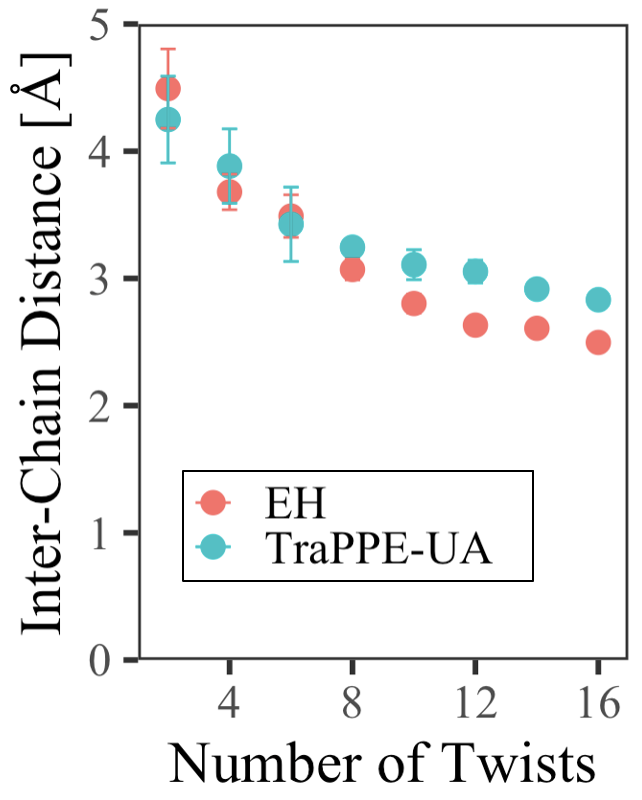}
\caption{Inter-chain distance between the chains in the double-helical polyethylene wire as a function of twist. Inter-chain distance is obtained by averaging several distances between corresponding atoms on opposite chains. Results are shown for (red) the Explicit Hydrogen model and (blue) the TraPPE-UA model. Standard deviations are shown when larger than symbol size.}
\label{FIG8}
\end{figure*}

Due to the dependence of axial angular momentum and the axial momentum pseudoscalar on the perpendicular radius, we estimate the average interchain distance as a function of number of twists for our double-helical polymer. We see that in both the TraPPE-UA and EH models, the inter-chain distance decreases monotonically with the number number of twists. Other physical changes induced by twist are discussed in Ref. \cite{twist_paper}.

\newpage
\pagebreak
\newpage

\bibliographystyle{aipnum4-1}

\end{document}